\documentclass[twocolumn,showpacs,preprintnumbers,amsmath,amssymb]{revtex4}

\usepackage{graphicx}
\usepackage{dcolumn}
\usepackage{bm}

\begin{document}

\preprint{Received 16 January 2004; published 28 May 2004 
          ~~~~~~~~~~~~~~~~~~~~~~~~~~~~~~~~~~~~~~~~~
         {\sl Physical Review Letters} {\bf 92}, 211102 (2004)}

\title{Deflagrations and Detonations in Thermonuclear Supernovae}

\author
{Vadim N. Gamezo,$^{1}$ Alexei M. Khokhlov,$^{2}$ Elaine S. Oran$^{1}$ }  
\affiliation{
\centerline{\hbox{~}}
\centerline{\hbox{
$^{1}$Laboratory for Computational Physics and Fluid Dynamics,
Naval Research Laboratory, Washington, D.C. 20375, USA
}}   
\centerline{\hbox{
$^{2}$Department of Astronomy and Astrophysics, University of Chicago,
Chicago, IL 60637, USA
}}
}

\begin{abstract}

We study a type Ia supernova explosion using three-dimensional
numerical simulations based on reactive fluid dynamics.  We consider a
delayed-detonation model that assumes a deflagration-to-detonation
transition.  In contrast to the pure deflagration model, the
delayed-detonation model releases enough energy to account for a
healthy explosion, and does {\it not} leave carbon, oxygen, and
intermediate-mass elements in central parts of a white dwarf. This
removes the key disagreement between simulations and observations, and
makes a delayed detonation the mostly likely mechanism for type Ia
supernovae.

\end{abstract}


\maketitle

Type Ia supernovae (SN~Ia) are produced by thermonuclear explosions of
white dwarf (WD) stars composed primarily of C and O nuclei and
detached degenerate electrons. The most probable explosion scenario
involves a binary star system, in which a WD can increase its own mass
by accreting material from its companion until it approaches the
Chandrasekhar limit, $1.4M_\odot$. Near this limit, a small increase in
mass results in substantial contraction and compression of the WD. The
compression increases the temperature, accelerates nuclear fusion
reactions, and triggers the thermonuclear runaway [1] that eventually
ignites thermonuclear burning near the WD center.  This starts a
thermonuclear explosion that releases $\sim$$10^{51}$ ergs during a few
seconds. The energy is produced by a network of thermonuclear reactions
that begins with the original $^{12}$C and $^{16}$O nuclei and ends in
the formation of $^{56}$Ni and other iron-group elements.  Considerable
amounts of intermediate-mass elements (IME), such as Ne, Mg, Si, S, and
Ca, are created as well. Type Ia and other types of supernovae play an
important role in stellar nucleosynthesis and in releasing the newly
synthesized elements into interstellar medium, thus providing raw
material for next generations of stars and planets.

Because of their extreme and predictable luminosity, SN~Ia are extensively
used as standard candles to measure distances and estimate cosmological
parameters critical for our understanding of the global evolution of
the universe. To improve these measurements, we need comprehensive
theoretical and numerical models of SN~Ia that describe details of the
explosion and connect them to observed characteristics of SN~Ia, such
as spectra and light curves.  One-dimensional (1D) numerical models
have been extensively used to test general ideas about possible
explosion mechanisms  [2-7]. In particular, 1D models have ruled out
the possibility of a thermonuclear detonation, a supersonic
shock-induced combustion mode, as a sole mechanism for SN~Ia
explosions.  A detonation propagating through a high-density WD
produces mostly Ni and almost none of IME observed in SN~Ia spectra.
One-dimensional models have also shown that a detonation can produce
IME if it propagates through a low-density WD preexpanded during the
initial deflagration stage of the explosion.  These delayed-detonation
models  [8-14], which have a deflagration-to-detonation transition
(DDT) at some stage of the thermonuclear explosion, are the most
successful in reproducing observed characteristics of SNe~Ia. Many
crucial physical details, however, including the mechanism of DDT and
the turbulent flame structure, are missing by definition from 1D models
because SN~Ia explosions are intrinsically three-dimensional (3D)
phenomena.

Full-scale 3D numerical simulations of thermonuclear supernova
explosions have become a reality during the past few years  [15-18].
They have shown, in particular, that the development of a turbulent
thermonuclear flame in the gravitational field of a WD allows funnels
of unburned and partially burned material to remain in the vicinity of
the WD center until the end of the explosion. This would produce
distinct signatures of low-velocity C, O, and IME in SN~Ia spectra.  As
the observed spectra do not show these signatures, the deflagration
model must be incomplete. Previously we suggested  [15,18]  but did not
prove that a detonation triggered by the turbulent flame could burn the
remaining material near the WD center and make the model consistent
with observations. Here, we test this hypothesis using a 3D numerical
delayed-detonation model of SN~Ia explosion in which a deflagration
undergoes a transition to a detonation.

{\it Input Physics and Numerical Implementation. ---} 
The numerical model discused in details in   [15,18] is based on
reactive Euler equations that include gravity terms and are coupled
with an equation of state for a degenerate matter and a simplified
kinetics of energy release. The equations are integrated on a Cartesian
adaptive mesh using an explicit, second-order, Godunov-type numerical
scheme. The model describes compressible fluid dynamics on large scales
in an exploding WD including the expansion of a star, Rayleigh-Taylor
(RT) and Kelvin-Helmholtz (KH) instabilities, turbulence, pressure
waves, shocks, and detonations.  The nuclear kinetics is approximated
by a four-equation mechanism  [8,15] that describes the energy release,
consumption of C, nuclear statistical quasi-equilibrium (NSQE or QSE)
and nuclear statistical equilibrium (NSE) relaxations, and
neutronization.  The turbulent flame speed is provided by an additional
subgrid model [15,18]  that takes into account physical processes at
scales smaller than the computational cell size.  In particular, it
assumes that turbulent burning on small unresolved scales is driven by
the gravity-induced RT instability.

The model is able to reproduce the two different regimes of the
thermonuclear burning in a WD, a subsonic deflagration and a supersonic
detonation.  These regimes differ by the mechanism of propagation of
the reaction wave: a deflagration involves heat conduction or turbulent
mixing, and a detonation involves shock compression. For both regimes,
the energy is released by the same network of thermonuclear reactions,
and the physical thickness of the reaction front strongly depends on
density.  It can be up to 12 orders of magnitude less than the WD
radius $R_{WD}$ for deflagrations  [19,20]  and up to 10 orders of
magnitude less than $R_{WD}$ for detonations  [21,22].  Since the
large-scale simulations described here do not resolve length scales
smaller than $10^{-3}R_{WD}$, the reaction fronts at high densities are
still unresolved. We explicitly resolve only parts of the reaction zone
associated with NSE relaxation that become very large at low densities
and cause an incomplete burning that produces Si and other IME.
Resolution tests show that the minimum computational cell size
$dx_{min} = 2.6\times10^5$~cm used here for the deflagration stage and
$dx_{min} = 10.4\times10^5$~cm used for the detonation stage are
adequate for this type of simulations.

{\it Deflagration Stage. ---}
The initial conditions for the deflagration stage model a $1.4M_\odot$
WD in hydrostatic equilibrium with initial radius
$R_{WD}=2\times10^8$~cm, initial central density $\rho_c=
2\times10^9$~g/cm$^3$, spatially uniform initial temperature
$T=10^5$~K, and uniform initial composition with equal mass fractions
of $^{12}$C and $^{16}$O nuclei. The burning was initiated at the
center of WD by filling a small spherical region at $r<0.015R_{WD}$
with hot reaction products without disturbing the hydrostatic
equilibrium.  We model one octant of the WD assuming mirror symmetry
along the $x=0$, $y=0$ and $z=0$ planes. The computational domain is a
cube with a side of $x_{max}=5.35\times10^8$~cm.

The development of the thermonuclear flame was described in detail
earlier  [18]. The initially spherical flame propagates outwards with a
laminar velocity $\sim100$~km/s, becomes distorted due to the RT
instability, and forms multiple plumes at different scales.  Buoyancy
causes the hot, burned, low-density material inside the flame plumes to
rise towards the WD surface. The same gravitational forces also pull
the cold, unburned, high-density material between the plumes down
towards the center. The flame becomes turbulent and forms a dynamic
convoluted surface penetrated in all directions by very irregular
funnels of unburned and partially burned material.

The higly developed 3D flame surface increases the burning rate and
improves the estimation of energy release compared to 1D and 2D
deflagration models.  The intense convection on large scales also
causes the burned material to spend less time in high-density central
parts of WD, thus reducing the neutronization [15]. The same convective
flows, however, bring unburned material to the central parts of the
exploding star.  As a result, substantial amounts of C, O, and IME
remain near the WD center by the end of the explosion.  As we have
shown  [15,18], this makes predictions of the 3D deflagration model
inconsistent with observed spectra of SN~Ia.

{\it Detonation Stage. ---}
The disagreement between predictions from the pure deflagration
simulation and observations strongly suggests that the turbulent flame
in SN~Ia triggers a detonation.  The process of DDT involves events
occurring at small scales that are comparable to the detonation wave
thickness, and, thus, cannot be directly modeled in large-scale
simulations.  To study the effects of a detonation, we therefore assume
a time and a location for DDT. (A similar approach has been used
previously in 1D  [8-13] and 2D  [23-25]  delayed-detonation models.)
Now the deflagration results are initial conditions, and we impose a
hot spot to ignite the detonation.  The time and location for the
detonation initiation are parameters that can be varied and optimized.
Here, we explore the three cases (a), (b), and (c) defined below.

The case (a) corresponds to central detonation initiation at 1.62~s
after the beginning of the deflagration. By that time, 1/3 of WD mass
has burned, the WD radius has increased by a factor of 1.55, and the
density of unburnt material near the center has dropped to
$2.5\times10^8$~g/cm$^3$.  A detonation at this density produces mostly
Ni and propagates outwards at $\sim$$12,000$~km/s, which is comparable
to the expansion velocities induced by subsonic burning. When the
detonation reaches unburned material with density below
$(1-5)\times10^7$~g/cm$^3$, it begins to produce IME.  Different parts
of the detonation front that exit different funnels collide with each
other, coalesce, and eventually reach the surface of the star.

\begin{figure}
\resizebox{70mm}{!} {\includegraphics{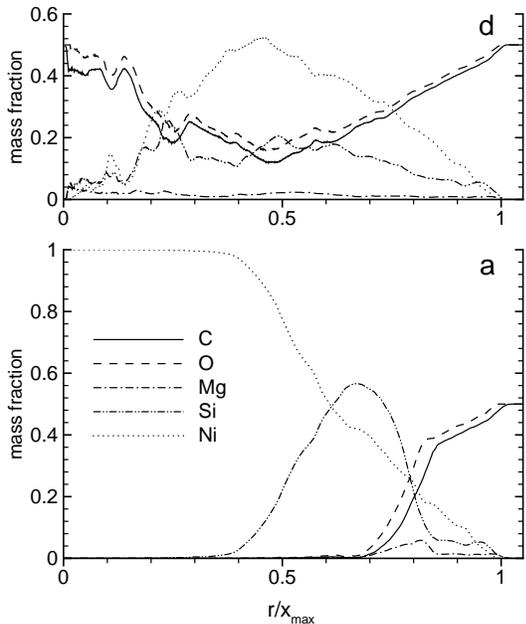}}
\caption{\label{fig:fig1}
Angle-averaged mass fractions of the main elements as functions of
scaled distance from the WD center produced by the deflagration (d) and
delayed-detonation (a) models at 1.94~s after the beginning of the
explosion. The delayed-detonation model corresponds to case (a)
described in the text.  $x_{max}=5.35\times10^8$~cm.  Lines marked as
Mg, Si, and Ni represent estimated cumulative mass fractions of
elements from Ne to Mg, Si to Ca, and Ti to Ni, respectively.  The
estimations are based on a four-equation nuclear kinetic scheme
[15,18] and  the reaction zone structure of a 1D detonation wave
calculated in [21] with a detailed nuclear kinetics.  
} \end{figure}

The detonation transforms all C and O in central parts of the WD into
iron-group elements, and produces IME in outer layers. This drastically
changes the distribution of nuclei compared to that produced by  the
pure deflagration.  Funnels of unburned C and O disappear from central
parts of the WD.  Iron-group elements form a distinct core surrounded
by a layer of IME. Angle-averaged mass fractions of the main elements
calculated for the deflagration and the delayed-detonation models are
compared in Fig.~\ref{fig:fig1}.

Similar results were obtained for the delayed-detonation case (b), for
which the detonation was initiated at 1.62~s at $10^8$~cm off center
and produced a moderate asymmetry in composition.  The results indicate
that, during the period of detonation propagation, the density of the
expanding unreacted material ahead of the shock can decrease by an
order of magnitude compared to its value at the end of the deflagration
stage. Because the detonation burns material to different products at
different densities, it can create a large-scale asymmetry in
composition if it starts far from the WD center. A similar conclusion
based on 2D simulations was made in  [25].

The asymmetry effect in our simulations is limited because we calculate
only one octant of a WD and impose mirror boundary conditions. The
degree of asymmetry would increase if the simulations were performed
for a full star. Then the second mirror-reflected spot for detonation
initiation would be eliminated.  Three-dimensional simulations [15,26]
also show that a developing flame, unrestricted by mirror boundaries,
can move away from the WD center, thus creating a large-scale asymmetry
at very early stages of the explosion.

\begin{figure}
\resizebox{65mm}{!} {\includegraphics{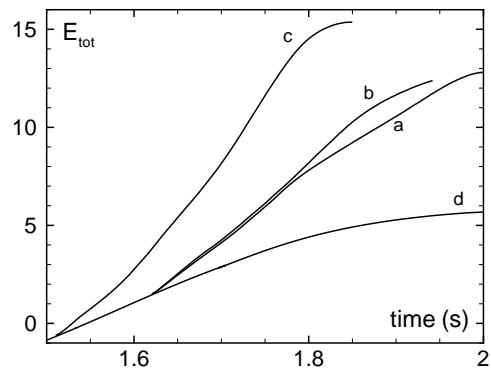}}
\caption{\label{fig:fig2}
Total energy as function of time for deflagration (d) and
delayed-detonation cases (a), (b), and (c) described in the text.
Energy units are $10^{50}$~ergs.
}
\end{figure}

For the case (c), the detonation was initiated at the WD center at
1.51~s when 1/4 of WD mass has burned, the WD radius has increased by a
factor of 1.30, and the density of unburnt material near the center has
decreased to $4.4\times10^8$~g/cm$^3$. Case (c) produced more
iron-group elements than cases (a) and (b) because the detonation
propagated through higher-density material. The earlier detonation
initiation also resulted in a faster explosion that released 15\% more
energy.  Total energies for all three cases and the deflagration model
are shown in Fig.~\ref{fig:fig2} as functions of time. The total energy
$E_{tot}$ here is the difference between the energy released by
thermonuclear reactions and the binding energy of the star. Eventually
$E_{tot}$ will be transformed into kinetic energy of expanding material
that can be measured in observations of SN~Ia.

Figure~\ref{fig:fig2} shows that the total energy released by
delayed-detonation models, $(1.3-1.6)\times10^{51}$~ergs, is much
higher than the energy released by the deflagration model
$\sim$$0.6\times10^{51}$~ergs. The reason for this is that the
deflagration is able to burn only about a half of the WD mass. The rest
of the material expands to the densities below $\simeq$$10^6$~g/cm$^3$
that do not support the thermonuclear burning. A detonation propagates
faster and burns almost all of the WD mass before the material expands
to low densities.  The total energy released by the delayed-detonation
models is in agreement with a typical range $(1-1.5)\times10^{51}$~ergs
obtained from SN~Ia observations  [7].

{\it Discussion and Conclusions. ---}
Figure~\ref{fig:fig1}  shows the averaged distribution of elements at
1.94~s, the time when the WD surface reached the computational domain
boundary, but the detonation did not yet reach the WD surface. At this
time, the detonation front propagates through low-density outer layers
of the star and produces mostly IME. All
iron-group elements have already formed at higher densities. The total
mass of iron-group elements created by the explosion is 0.78, 0.73, and
0.94 solar masses ($M_\odot$) for delayed-detonation cases (a), (b),
and (c), respectively. Most of this mass is the radioactive $^{56}$Ni
that provides the energy source for the observed luminosity of SN~Ia.
The mass of $^{56}$Ni estimated from observational data is about
$0.6M_\odot$ for a typical SN~Ia  [27], and is in agreement with the
total mass of iron-group elements produced by delayed-detonation
models. For the deflagration model, the total mass of iron-group
elements is only 0.47 $M_\odot$, which is insufficient to account for
the luminosity of a typical SN~Ia.

The carbon-oxygen layer that remains between the detonation front and
the WD surface will continue to burn as the detonation advances. Oxygen
in outer layers, which expand to densities below
$\simeq$$10^6$~g/cm$^3$ before the detonation reaches them, will remain
unburned. Carbon is likely to remain unburned for densities below
$(1-3)\times10^5$~g/cm$^3$. Unburned C and O in outer layers
would produce spectral signatures only in the high-velocity range.

There are possibilities, however, for a delayed detonation to leave
small amounts of C and O in inner parts of WD. For example, a
detonation propagating through a thin, sinuous funnel of unburned
material can fail if the funnel makes a sharp turn. A developing
turbulent flame can also disconnect some funnels from the rest of the
unburned material, thus creating unburned pockets that cannot be
directly reached by a detonation wave. These pockets may or may not
ignite when strong shocks generated by detonations reach them.  The
cellular structure of thermonuclear detonations in carbon-oxygen
matter  [22], and the ability of cellular detonations to form pockets
of unburned material that extend far behind the 1D reaction zone, can
also contribute into incomplete burning. All these phenomena occur at
length scales comparable to the reaction zone thickness that are not
resolved in large-scale simulations reported here, and require
additional studies.

There have recently been efforts to detect low-velocity C in SN~Ia
spectra that could result from the funnels of unburned material near
the WD center  [28]. The results  [28]  indicate that C can be
present at velocities 11~000~km/s.  Even though this velocity is much
lower than 20~000-30~000~km/s usually attributed to C in SN~Ia
spectra  [29-31], it is still too high for the ejecta formed from
central parts of a WD. For C and O, spectral signatures are
difficult to observe, and estimated velocities of these elements are
subject to large uncertainties.  Intermediate-mass elements, however,
produce distinct spectral lines and their velocities are well defined.
The minimum observed velocities for IME
[28,32]  are large enough ($\sim10~000$~km/s for Si) to rule out the
presence of IME near the WD center, as is predicted by the
deflagration model.  A discussion on this subject can also be found in
the recent article  [33].

Figure~\ref{fig:fig1} shows that, in contrast to the 3D deflagration
model, the 3D delayed-detonation model of SN~Ia explosion does {\it
not} leave C, O, and IME in central
parts of a WD. This removes the key disagreement between simulations
and observations, and makes the 3D delayed detonation a promising
mechanism for SN~Ia explosion.  Further analysis of 3D delayed
detonations on large scale requires 3D radiation transport simulations
to produce spectra, and a detailed comparison between the calculated
and observed spectra of SN~Ia for different initiation times and
locations. The uncertainty in detonation initiation can only be
eliminated by solving the DDT problem that involves physical processes
at small scales.

This work was supported in part by the NASA ATP program
(NRA-02-OSS-01-ATP) and by the Naval Research Laboratory (NRL) through
the Office of Naval Research. Computing facilities were provided by the
DOD HPCMP program. We would like to thank Peter H\"oflich and Craig Wheeler
for helpful discussions.

\vfill

\newcounter{Lcount}
\begin{list} 
   {[\arabic{Lcount}]} 
   {\usecounter{Lcount} \itemsep=0pt \parsep=0pt \leftmargin=5mm }

\item 
F.~Hoyle, W.~A.~Fowler, {\sl Astrophys. J.} {\bf 132}, 565 (1960)

\item 
W.~D.~Arnett, {\sl Astrophys. Space Sci.} {\bf 5}, 180 (1969)

\item 
C.~J.~Hansen, J.~C.~Wheeler, {\sl Astrophys. Space Sci.} {\bf 3}, 464 (1969)

\item 
K.~Nomoto, D.~Sugimoto, S.~Neo, {\sl Astrophys. Space Sci.} {\bf 39}, L37 (1976)

\item 
K.~Nomoto, F.-K.~Thielemann, K.~Yokoi, {\sl Astrophys. J.} {\bf 286}, 644 (1984)

\item 
S.~E.~Woosley, T.~A.~Weaver, {\sl Annu. Rev. Astron. Astrophys.} {\bf 24},
205 (1986)

\item 
J.~C.~Wheeler, R.~P.~Harkness, A.~M.~Khokhlov, P.~A.~H\"oflich,
{\sl Phys. Rep.} {\bf 256}, 211 (1995)

\item 
A.~M.~Khokhlov, {\sl Astron. Astrophys.} {\bf 245}, 114 (1991)

\item 
H.~Yamaoka, K.~Nomoto, T.~Shigeyama, F.-K.~Thielemann,
{\sl Astrophys. J.} {\bf 393}, L55 (1992)

\item 
A.~M.~Khokhlov, E.~M\"uller, P.~A.~H\"oflich, {\sl Astron. Astrophys.}
{\bf 270}, 223 (1993)

\item 
P.~A.~H\"oflich, {\sl Astrophys. J.} {\bf 443}, 89 (1995)

\item 
P.~A.~H\"oflich, A.~M.~Khokhlov, J.~C.~Wheeler,
{\sl Astrophys. J.} {\bf 444}, 831 (1995)

\item 
P.~A.~H\"oflich, A.~M.~Khokhlov, {\sl Astrophys. J.} {\bf 457}, 500 (1996)

\item 
J.~C.~Niemeyer, S.~E.~Woosley, {\sl Astrophys. J.} {\bf 475}, 740 (1997)

\item 
A.~M.~Khokhlov, astro-ph/0008463 (2000)

\item 
M.~Reinecke, W.~Hillebrandt, J.~C.~Niemeyer, {\sl Astron. Astrophys.}
{\bf 386}, 936 (2002)

\item 
M.~Reinecke, W.~Hillebrandt, J.~C.~Niemeyer, {\sl Astron. Astrophys.},
{\bf 391}, 1167 (2002)

\item 
V.~N.~Gamezo,  A.~M.~Khokhlov, E.~S.~Oran, A.~Y.~Chtchelkanova, 
R.~O.~Rosenberg, {\sl Science}, 299, 77 (2003)

\item 
F.~X.~Timmes, S.~E.~Woosley, {\sl Astrophys. J.} {\bf 396}, 649 (1992)

\item 
A.~M.~Khokhlov, E.~S.~Oran, J.~C.~Wheeler, {\sl Astrophys. J.}
{\bf 478}, 678 (1997)

\item 
A.~M.~Khokhlov, {\sl Mon. Not. R. Astron. Soc. } {\bf 239}, 785 (1989)

\item 
V.~N.~Gamezo, J.~C.~Wheeler, ~A.~M.~Khokhlov, E.~S.~Oran,
{\sl Astrophys. J.} {\bf 512}, 827 (1999)

\item 
D.~Arnett, E.~Livne, {\sl Astrophys. J.} {\bf 427}, 315 (1994)

\item 
D.~Arnett, E.~Livne, {\sl Astrophys. J.} {\bf 427}, 330 (1994)

\item 
E.~Livne, {\sl Astrophys. J.} {\bf 527}, L97 (1999)

\item 
A.~C.~Calder {\sl et al}., astro-ph/0405162.

\item 
D.~Branch, A.~M.~Khokhlov, {\sl Phys. Rep.} {\bf 256}, 53 (1995)

\item 
D.~Branch {\sl et al}., {\sl Astron. J.} {\bf 126}, 1489 (2003)

\item 
R.~P.~Kirshner {\sl et al}., {\sl Astrophys. J.} {\bf 415}, 589 (1993)

\item 
A.~Fisher, D.~Branch, P.~Nugent, E.~Baron, {\sl Astrophys. J.} {\bf 481}, L89 (1997)

\item 
P.~A.~Mazzali, {\sl Mon. Not. R. Astron. Soc. } {\bf 321}, 341 (2001)

\item 
A.~V.~Filippenko, {\sl Annu. Rev. Astron. Astrophys.} {\bf 35}, 309 (1997)

\item 
D.~Branch, astro-ph/0310685.

\end{list}

\end{document}